\input{aipcheck}

\documentclass[
    ,final            
  ]
  {aipproc}

\layoutstyle{6x9}

\usepackage{amsmath}
\usepackage{graphics}
\usepackage{graphicx}
\usepackage[mathcal]{eucal}

\usepackage{epsfig}
\usepackage{dcolumn}
\usepackage{bm}

\begin{document}

\title{A Viable Flavor Model for Quarks and Leptons in RS with $T^{\prime}$ Family Symmetry}


\author{Mu-Chun Chen}{
  address={Department of Physics \& Astronomy, University of California, Irvine, CA 92697-4575, USA}
}

\author{K.T. Mahanthappa}{
  address={Department of Physics, University of Colorado, Boulder, CO 80309-0390, USA}
}

\author{Felix Yu}{
  address={Department of Physics \& Astronomy, University of California, Irvine, CA 92697-4575, USA}
}

\begin{abstract}
We propose a Randall-Sundrum model with a bulk family symmetry based
on the double tetrahedral group, $T^{\prime}$, which generates the
tri-bimaximal neutrino mixing pattern and a realistic CKM matrix.   
The $T^{\prime}$ symmetry forbids tree-level
flavor-changing-neutral-currents in both the quark and lepton sectors,
as different generations of fermions are unified into multiplets of
$T^{\prime}$. This results in a low first KK mass scale and thus the
model can be tested at collider experiments.
\end{abstract}

\maketitle


\section{Introduction}
\label{intro}

The Randall-Sundrum (RS) Model~\cite{Randall:1999ee} has been proposed as a
non-supersymmetry alternative solution to the gauge hierarchy
problem. In addition to solving the gauge hierarchy problem, the model
can accommodate the fermion mass hierarchy, by localizing different fermions at different
points in the fifth dimension~\cite{bulkfermion1,bulkfermion2}. 
The presence of the 5D bulk mass parameters generically lead to dangerously large
flavor-changing-neutral-currents (FCNCs) already at the tree level
through the exchange of the KK gauge bosons. One way to avoid the tree level FCNCs is by imposing minimal flavor
violation~\cite{mfv} which assumes that all flavor violation
comes from the Yukawa sector~\cite{Fitzpatrick:2007sa,Chen:2008qg}. 
We propose~\cite{Chen:2009gy} an alternative by imposing a bulk family
symmetry \cite{Csaki:2008qq} based on the double tetrahedral
group~\cite{Frampton:1994rk,Chen:2007afa,Chen:2009gf},
$T^{\prime}$, which 
simultaneously gives rise to tri-bimaximal (TBM) neutrino mixing and a realistic
CKM matrix. In addition, the complex CG coefficients of $T^{\prime}$ also give the possibility that CP
violation is entirely geometrical in origin~\cite{Chen:2009gf}. Because the three lepton
doublets form a $T^{\prime}$ triplet while the
three generations of quarks transform as $2 \oplus 1$, these multiplets now have common bulk mass terms. 
This assignment thus forbids tree-level FCNCs involving the first and
second generations of quarks, which are the most severely constrained, as well as all tree-level FCNCs in the lepton sector.

\section{The Model}

The particle content of our model is given in Ref.~\cite{Chen:2009gy}. The 5D Lagrangian involving leptons is
$\mathcal{L}^{\text{lep}}_{\text{5D}} \supset
\mathcal{L}^{\text{lep}}_{\text{Kin}} +
\mathcal{L}^{\text{lep}}_{\text{Bulk}} +
\mathcal{L}^{\text{lep}}_{\text{Yuk, } \ell} +
\mathcal{L}^{\text{lep}}_{\text{Yuk, } \nu}$, where the bulk mass terms are
\begin{equation}
\mathcal{L}^{\text{lep}}_{\text{Bulk}} = k \left( \overline{L} c_L L +
\overline{e} c_e e + \overline{\mu} c_{\mu} \mu + \overline{\tau}
c_{\tau} \tau + \overline{N} c_N N \right) \ ,
\label{eqn:Llepmass}
\end{equation}
and the 5D Yukawa interactions for charged leptons and neutrinos are
\begin{eqnarray}
\mathcal{L}^{\text{lep}}_{\text{Yuk, } \ell} & = & \delta 
\left( y - \pi R \right) \left[ \frac{1}{k} \overline{H}(x)
\left( y_e^{\text{5D}} \overline{L}(x,y) e(x,y) \frac{\phi(x)}{\Lambda}
\right.\right.\nonumber \\
& & 
\left.\left.+ y_{\mu}^{\text{5D}} \overline{L}(x,y) \mu(x,y) \frac{\phi(x)}{\Lambda}
+y_{\tau}^{\text{5D}} \overline{L}(x,y) \tau(x,y) 
\frac{\phi(x)}{\Lambda} \right) \right] + \text{ h.c.}
\label{eqn:Llepyuk}
\end{eqnarray}
\begin{eqnarray}
\mathcal{L}^{\text{lep}}_{\text{Yuk, } \nu, \text{ SS}} & = & 
\left\{ \delta \left( y \right) \left[ \frac{1}{k} N^T (x,y) N(x,y) 
\left( y_{\nu, \text{ SS, a}}^{\text{5D}} 
\phi_{\text{SS}}^{\prime}(x') 
+ y_{\nu, \text{ SS, b}}^{\text{5D}} \sigma_{\text{SS}}(x') 
\right) \right] \right. \nonumber \\
& & \left. 
+ \delta \left( y - \pi R \right) \left[ \frac{1}{k} H(x) 
y_{\nu, \text{ SS, c}}^{\text{5D}} \overline{L}(x,y) N(x,y) 
\right] \right\} + \text{ h.c.} \ ,
\label{eqn:Lnuyuk2}
\end{eqnarray}
The flavons fields acquire the following VEVs,
\begin{equation}
\langle \phi \rangle = \phi_0 \Lambda \left( \begin{array}{c} 1 \\ 0
\\ 0 \\
\end{array} \right) \qquad 
\langle \phi_{\text{SS}}^{\prime} \rangle = 
\phi_{0, \text{ SS}}^{\prime} \Lambda_{\text{UV}} \left(
\begin{array}{c} 1 \\ 1 \\ 1 \\ \end{array} \right) \qquad
\langle \sigma_{\text{SS}} \rangle = \sigma_{0, \text{ SS}}
\Lambda_{\text{UV}} \ ,
\label{eqn:lepvev1}
\end{equation}
leading to charged lepton mass matrix,
\begin{equation}
M_e = v \phi_0 \left( \begin{array}{ccc}
y_e & 0 & 0 \\
0 & y_\mu & 0 \\
0 & 0 & y_\tau \\
\end{array} \right) \ , \quad y_\ell = y_\ell^{\text{5D}} f(c_L, c_\ell) \; .
\label{eqn:Lmatrix}
\end{equation}
The 4D effective Majorana mass matrix for RH neutrinos is
\begin{equation}
M_{\text{RR}} = \frac{1 - 2 c_N}{2 \left( e^{(1 - 2 c_N) \pi k R} 
- 1 \right)} \Lambda_{\text{UV}}
\left( \begin{array}{ccc}
2A + B & -A & -A \\
-A & 2A & B - A \\
-A & B - A & 2A \\
\end{array} \right) \ ,
\label{eqn:Nmatrix2a}
\end{equation}
where 
\begin{equation}
A = \frac{1}{3} y_{\nu, \text{ SS, a}}^{\text{5D}} 
\phi_{0, \text{ SS}}^{\prime} \qquad
B = y_{\nu, \text{ SS, b}}^{\text{5D}} \sigma_{0, \text{ SS}} \ ,
\end{equation}
and the neutrino Dirac mass matrix is
\begin{equation}
M_{\text{Dc}} = y_{\nu, {\text{ SS, c}}}^{\text{5D}} f(c_L, c_N) v
\left( \begin{array}{ccc}
1 & 0 & 0 \\
0 & 0 & 1 \\
0 & 1 & 0 \\
\end{array} \right) \ .
\label{eqn:Nmatrix2b}
\end{equation}
The resulting light effective LH neutrino
mass matrix is form diagonalizable~\cite{Chen:2009um} by the TBM mixing matrix to
give eigenvalues
\begin{equation}
M_{\nu, \text{ eff}}^D = 
\frac{ \left( y_{\nu, \text{ SS, c}}^{\text{5D}}
 v \right)^2 }{2 \Lambda_{\text{UV}} } 
\frac{ 1 - 2 c_L}{e^{(1 - 2 c_L) \pi k R} - 1}
e^{2 (1 - c_L - c_N) \pi k R} \text{ diag} \left( \frac{1}{3A + B},
\frac{1}{B}, \frac{1}{3A - B} \right) \ .
\end{equation}

Due to universal $c_L$ for the lepton doublets and RH neutrinos, all tree-level leptonic FCNCs are absent. 
The observed charged lepton mass hierarchy can be obtained via the non-universal values for
the bulk mass parameters for the RH charged leptons, $c_\ell$.

The 5D Lagrangian involving quarks is given by, 
$\mathcal{L}^{\text{qrk}}_{\text{5D}} \supset
\mathcal{L}^{\text{qrk}}_{\text{Kin}} +
\mathcal{L}^{\text{qrk}}_{\text{Bulk}} +
\mathcal{L}^{\text{qrk}}_{\text{Yuk}}$, where the bulk mass terms are
\begin{equation}
\mathcal{L}^{\text{qrk}}_{\text{Bulk}} = k \left( \overline{Q}_{12}
c_{Q_{12}} Q_{12} + \overline{Q}_3 c_{Q_3} Q_3 + \overline{U} c_U U +
\overline{T} c_T T + \overline{D} c_D D + \overline{B} c_B B \right)
\ ,
\label{eqn:Lqrkmass}
\end{equation}
and the 5D Yukawa interactions are 
\begin{eqnarray}
\mathcal{L}^{\text{qrk}}_{\text{Yuk}} &=& 
 \left\{ \frac{1}{k} H(x) \left[
  y_{12}^U \overline{Q}_{12}(x,y) U(x,y) 
          \left(\frac{\alpha(x) + \zeta(x)}{\Lambda} \right) +
  y_3^U \overline{Q}_3(x,y) U(x,y) \frac{\chi_U(x)}{\Lambda} 
\right. \right. \nonumber \\
& & \left. + 
  y_{12}^T \overline{Q}_{12}(x,y) T(x,y) \frac{\eta_U(x)}{\Lambda} +
  y_3^T \overline{Q}_3(x,y) T(x,y) \right] \nonumber \\
& & + \frac{1}{k} \overline{H}(x) \left[
  y_{12}^D \overline{Q}_{12}(x,y) D(x,y) 
          \left(\frac{\beta(x) + \xi(x)}{\Lambda} \right) +
  y_3^D \overline{Q}_3(x,y) D(x,y) \frac{\chi_D(x)}{\Lambda} 
\right. \nonumber \\ 
& & \left. \left. + 
  y_{12}^B \overline{Q}_{12}(x,y) B(x,y) \frac{\eta_D(x)}{\Lambda} +
  y_3^B \overline{Q}_3(x,y) B(x,y) \right] \right\} \delta \left( y - \pi R \right)  \ .
\label{eqn:Lqrkyuk}
\end{eqnarray}
The flavons acquire VEVs along the following directions,
\begin{equation}
\langle \alpha \rangle = \langle \beta \rangle = \Lambda \alpha 
\left( \begin{array}{c} 
1 \\ 1 \\ 1 \\
\end{array} \right), \; 
\langle \zeta \rangle = \langle \xi \rangle = \Lambda \zeta, \;
\langle \chi_r \rangle = \Lambda \chi_r \left( \begin{array}{c}
\cos \theta_r \\ \sin \theta_r \\
\end{array} \right), \; 
\langle \eta_r \rangle = \Lambda \eta_r \left( \begin{array}{c}
1 \\ 0 \\
\end{array} \right) \ ,
\label{eqn:qrkvev}
\end{equation}
where $r = U$, $D$.  This VEV pattern leads to a 4D effective up-type
mass matrix
\begin{equation}
\frac{M_U}{v} = \left( \begin{array}{ccc}
i \alpha f(c_{Q_{12}}, c_U) & 
\left[ \left( \frac{1-i}{2} \right) \alpha - \zeta \right] 
f(c_{Q_{12}}, c_U) 
& \chi_U \sin \theta_U f(c_{Q_3}, c_U) \\
\left[ \left( \frac{1-i}{2} \right) \alpha + \zeta \right] 
f(c_{Q_{12}}, c_U) 
& \alpha f(c_{Q_{12}}, c_U) & -\chi_U \cos \theta_U f(c_{Q_3}, c_U) \\
0 & -\eta_U f(c_{Q_{12}}, c_T) & y_3^T f(c_{Q_3}, c_T) \\
\end{array} \right)
\label{eqn:Umatrix}
\end{equation}
and a 4D effective down-type mass matrix
\begin{equation}
\frac{M_D}{v} =  \left( \begin{array}{ccc}
i \alpha f(c_{Q_{12}}, c_D) & 
\left[ \left( \frac{1-i}{2} \right) \alpha - \zeta \right] 
f(c_{Q_{12}}, c_D) 
& \chi_D \sin \theta_D f(c_{Q_3}, c_D) \\
\left[ \left( \frac{1-i}{2} \right) \alpha + \zeta \right] 
f(c_{Q_{12}}, c_D) 
& \alpha f(c_{Q_{12}}, c_D) & -\chi_D \cos \theta_D f(c_{Q_3}, c_D) \\
0 & -\eta_D f(c_{Q_{12}}, c_B) & y_3^B f(c_{Q_3}, c_B) \\
\end{array} \right) \ ,
\label{eqn:Dmatrix}
\end{equation}
where all other Yukawas have been set to 1.

Similar to the lepton sector, due to the $T^{\prime}$ family symmetry
and the $2 \oplus 1$ structure, the number of bulk parameters is
greatly reduced in our model, and the tree-level FCNCs involving the first and second generations of quarks are absent.

\section{Numerical Results}
The number of independent parameters for our model is much smaller compared to generic RS models: we
have 3 for the charged leptons, 2 for the neutrinos, and 11 for the
quarks, which compares to 30 for the anarchic leptons and 36 for
the anarchic quarks. With $c_L = 0.40000$, $c_e = 0.82925$, $c_{\mu} = 0.66496$, $c_{\tau} = 0.57126$,
and $\phi_0 = 1$ as our input parameters, all charged lepton masses are obtained. 
To produce a normal hierarchy in the neutrino sector, we use the value of $c_L$ above, $c_N = 0.40000$, $\phi_{0, \text{
SS}}^{\prime} = 0.07427$, $\sigma_{0, \text{ SS}} = 0.06191$, leading to $m_1 = 0.004465$ eV, $m_2 =
0.009821$ eV, and $m_3 = 0.04919$ eV.  An inverted hierarchy solution arises if we use $c_N
= 0.40000$, $\phi_{0, \text{ SS}}^{\prime} = 0.02321$, $\sigma_{0,
\text{ SS}} = -0.0115241$, which predict $m_1 = 0.05203$ eV, $m_2 = -0.05276$ eV, and
$m_3 = 0.01751$. Realistic quark masses and CKM mixing angles are obtained with 
 $\alpha = 0.0021i$, $\zeta = -0.0125$, $\chi_U = \eta_U = 0.52$, $\theta_U = -0.0825 \pi$, $\chi_D = 0.06615$,
$\theta_D = -0.0515 \pi$, $\eta_D = 0.2484 - 0.2024i$, $y_3^T =
0.7800$, and $y_3^B = 0.3000$,  $c_{Q_{12}}
= 0.435$, $c_{Q3} = 0.070$, $c_U = 0.535$, $c_T = -1.250$, $c_D =
0.562$, and $c_B = 0.521$.

\section{Conclusion}
We have proposed a Randall-Sundrum Model with a bulk $T^{\prime}$
family symmetry. The $T^{\prime}$ symmetry gives rise to a TBM mixing
matrix for the neutrinos and a realistic quark CKM matrix. Due to the
wavefunction localization, our model also naturally generates the
fermion mass hierarchy for both the quark and lepton sectors. The
$T^{\prime}$ representation assignments required for giving realistic
masses and mixing patterns automatically forbid all leptonic
tree-level FCNCs and those involving the first and the second
generations of quarks, which are present in generic RS models. As a
result, a low scale for the first KK scale mass can be allowed,
rendering the RS model a viable solution to the gauge hierarchy
problem and making it testable at collider experiments.


\begin{theacknowledgments}
 The work of M-CC was supported, in part, by the National Science Foundation under grant No. PHY-0709742. 
 The work of KTM was supported, in part, by the Department of Energy under grant No.  DE-FG02-04ER41290.
\end{theacknowledgments}


\end{document}